\documentstyle[twocolumn,prl,aps]{revtex}
\begin{document}
\title{Classical trajectories compatible with quantum mechanics}
\author{B. Roy Frieden}
\address{Optical Sciences Center, University of Arizona, Tucson AZ 85721 \\
A. Plastino \\
Physics Department, National University La Plata and \\
Argentine National Research Council (CONICET)\\
C.C. 727, 1900 La Plata, Argentina.}
\maketitle

\begin{abstract}
Consider any stationary Schroedinger wave equation (SWE) solution $\psi (x)$
for a particle. The corresponding PDF on position $x$ of the particle is $%
p_{X}(x)=\vert\psi (x)\vert^{2}$. There is a classical trajectory $x(t)$ for
the particle that is consistent with this PDF. The trajectory is unique to
within an additive constant corresponding to an initial condition $x(0)$.
However the value of $x(0)$ cannot be known. As an example, a free particle
in its ground state in a box of length $L$ obeys a classical trajectory $%
\,x/L-(1/2\pi )\sin (2\pi x/L)+t_{0}=t.$ The constant $t_{0}$ is an
unknowable time displacement. Momentum values, however, cannot be determined
by merely differentiating $d/dt$ the trajectory $x(t)$ and, instead, follow
the usual quantification rules of Heisenberg's. This permits position and
momentum to remain complementary variables. Our approach is fundamentally
different from that of D. Bohm. \vskip 2mm

PACS numbers: 03.65.Bz, 03.65.Sq, 05.30.-d, 5.40.-a
\end{abstract}

%%%%%%%%%%%%%%%%%%%%%%%%%%%%%%%%%%%%%%%%%%%%%%%%%%%%%%%%%%%%%%%%%%%%%%%%%%%%%%%%%%%%%%%%%%%%%%%%%%%%%%%%%%%%%%%%%%%%%%%%%%%%

\section{INTRODUCTION}

Suppose that a non-relativistic particle is allowed to move in one dimension 
$x$ (for simplicity) under the influence of a potential function. There are
two general perspectives for analyzing the motion of the particle: (i) that
of classical physics, according to which the particle obeys Newtonian
mechanics, or (ii) that of quantum mechanics, according to which the
particle obeys, e.g., the Schroedinger wave equation (SWE). (We ignore for
the moment the combined quantum-classical approach of D. Bohm [1].) Here we
will show that there is, in fact, {\it a third, intermediary viewpoint,}
whereby the particle obeys to a degree both classical mechanics and quantum
mechanics.

Let the particle have mass $m$, and be moving under the influence of a
stationary potential $V(x)$. Later on we treat the more general,
time-dependent case. The particle is otherwise not perturbed by any exterior
effect including measurement. Assuming non-relativistic speeds, the
kinematics of the particle obey the Schroedinger wave equation (SWE). In the
presence of our stationary potential this has as its solution a separable
wave amplitude function $\Psi (x,t)=\psi (x)\exp (iEt/\hbar )$, where $\psi
(x)$ is any complex amplitude function, $E$ is a definite energy value, and $%
\hbar $ stands for Planck's constant (divided by $2 \pi$).

The probability density function (PDF) for the separable case loses the
time-dependence, $p_{X}(x)=|\psi (x)|^{2}$. This means that, from a quantum
viewpoint, the coordinate $x$ is independent of the time $t$. The same PDF
on position is obeyed at all times. However, from a classical viewpoint, in
the presence of the potential $V(x)$ a particle has a {\em definite
trajectory }$x(t)$ through time. To what extent can the quantum result and
the classical viewpoint agree?

\section{TRAJECTORY}

Regard the particle as having a definite trajectory $x(t)$ during a time
interval $(0,T)$. Consider the histogram of position values $x$ taken on by
the particle during that time. The main point of this paper is that {\em for
any quantum} {\em solution} $\psi(x)$ there is a classical trajectory $x(t)$
whose density of position values $x$ over the given time interval exactly
coincides with $\vert \psi (x) \vert^{2}$. This trajectory is found next.

Central to the calculation is the following effect. Among the four
coordinates of space-time, each {\em time} coordinate value is unique in
occurring once and only once over a given interval $(0,T)$. That is, the
time values as coordinates occur predictably {\em in sequence} one after the
other. Of course, such a sequence of numbers has a {\em density} $p_{T}(t)$
of values that is uniform or flat over any time interval $(0,T),$

\begin{equation}
p_{T}(t)=1/T,\,\,t=(0,T).  \label{(1)}
\end{equation}
This expression is unusual in expressing a density of deterministic events.
However, whether deterministic or random, the density still must obey
Jacobian transformation theory [2].

One defining property of a trajectory $x(t)$ is that to each time value $t$
in the sequence there is one and only one $x$ value. Consider the density $%
p_{X}(x)$ of all $x$ values traveled over by the particle during the given
time interval. Note that like $p_{T}(t),$ density $p_{X}(x)$ is not a {\it %
probability} density, since the events $x$ follow a deterministic trajectory 
$x(t)$. Since to each $t$ value there is only one $x$ value the frequency of
occurrence of each $x$ (actually, of each interval $(x,x+dx)$) equals that
of its corresponding value $t$. Or, in terms of density functions [2],

\begin{equation}
p_{X}(x)|dx|=p_{T}(t)|dt|,  \label{(2)}
\end{equation}
where $x=x(t)$. Combining this with Eq. (1) gives

\begin{equation}
|\frac{dx}{dt}|=\frac{1}{T\,p_{X}(x)}.  \label{(3)}
\end{equation}
The absolute value sign for {\em dx/dt} defines two possible scenarios:
either {\em x} increases with {\em t} during the entire time period {\em T}
or {\em x} decreases with{\em \ t }during the period. (Note that for reasons
of continuity in {\em x}, the choice of sign cannot be changed {\em during}
the period.)\ By either sign,{\em \ x }is a monotonic function of {\em t}. \
Thus the particle makes one pass through the system in either the positive 
{\em x }direction or in the negative{\em \ x} direction. \ However, since
the trajectory obeys the required PDF $p_{X}(x)$ via Eq. (3) for {\em either
direction} (either choice of sign for {\em dx/dt}), the particle can
alternatively obey {\em periodic motion}. \ This would be for any number of
such periods {\em T}, i.e., any number of traversals to and fro through the
system. \ \ At each such traversal the particle position would be a
monotonic function of the time (as above) as defined by Eq. (3). \ In
summary, the particle motion can be either single pass, or oscillatory.

To be definite, consider the monotonically {\em increasing} orbit {\em %
dx(t)/dt} $\geq 0.$ Let the density $p_{X}(x)$ of {\it deterministic} events 
$x$ equal the density $|\psi (x)|^{2}$ of {\it random} events $x$ as given
by the SWE. \ Then Eq. (3) shows that the probabilistic density $|\psi
(x)|^{2}$ is achieved by a classical trajectory $x(t)$ obeying

\begin{equation}
\frac{dx}{dt}=\frac{1}{T\,\,|\psi (x)|^{2}}.  \label{(4)}
\end{equation}
This shows a possible problem at positions{\em \ x} for which $|\psi
(x)|^{2} $ $\rightarrow 0.$ \ The required particle speed {\em dx/dt}
becomes so large as to exceed {\em c}, the speed of light. \ However, Eq.
(4) also shows an inverse dependence upon the period {\em T,} and {\em T}
has yet to be chosen. \ Obviously, if $|\psi (x)|^{2}$ is small but not zero
the problem can be avoided by making {\em T} large enough. \ Furthermore,
even at {\em x} for which $|\psi (x)|^{2}$ $=0,$ with {\em T }chosen large
enough the unphysical region {\em x} within which relativity is violated can
be made as narrow (as improbable) as is required. \ 

Continuing with the analysis, Eq. (4) is straightforward to integrate, giving

\begin{equation}
\int dx|\psi (x)|^{2}=(t-t_{0})/T,  \label{(5)}
\end{equation}
with $t_{0}=\,\,const$. The integral is an indefinite one, resulting in a
well-defined function $g(x)$ corresponding to the given $\psi (x).$ Hence
Eq. (5) expresses the trajectory as

\begin{equation}
g(x)=(t-t_{0})/T,  \label{(6)}
\end{equation}
which has to be inverted to obtain the desired trajectory

\begin{equation}
x=g^{-1}[(t-t_{0})/T]\equiv x(t).  \label{(7)}
\end{equation}
This is the {\it formal} solution to the problem.

There are two distinct interpretations to this result. As it was derived, it
gives the trajectory of $x$ values that, if binned {\it deterministically}
at each time $t$ over the interval $(0,T)$ will give the required PDF $|\psi
(x)|^{2}$. But alternatively, the binned trajectory must also give the
required PDF if it is sampled with {\it uniform randomness} (1) in time. The
analysis is blind to whether $t$ is random or deterministic, since it
depends only upon the Jacobian transformation of density functions, and
Jacobian theory holds irrespective of whether the densities define
deterministic or random variables.

The time $t_{0}$ is of interest. With the function $g(x)$ fixed for a given
problem, Eq. (6) shows that the value of $t_{0}$ is defined by knowledge of
initial conditions. For example, if it is known that at $t=0$ position $x$
has a certain value $x_{0}$ then $t_{0}$ is fixed as $-T\,g(x_{0}) $.
Therefore, to know $t_{0}$ requires observation of the particle at a
particular space-time point. But if this observation were made the
particle's state would be changed, and it would no longer have the amplitude 
$\psi (x)$ that has been presumed in the calculation. In fact the state
would be a function of time, contrary to the stationary state solution
assumed.

Therefore the time $t_{0}$ remains an unknown constant of the theory, and
the trajectory (7) suffers an unknown time displacement. We conclude that 
{\it the trajectory is deterministic only up to an unknown additive constant}%
. Another way of saying this is that the trajectory is one of a family of
curves $x(t;t_{0})$ where each member of the family is defined by a
particular value of $t_{0}$. This defines, in fact, a stochastic process
[2]. We spoke at the beginning of defining a form of mechanics that is
intermediary between classical and quantum mechanics. It is this stochastic
process that is intermediary between the two, preserving aspects of each.
Thus, the trajectory $x(t)$ is deterministic, or classical, to an additive
random constant, and {\it it is the latter that gives a quantum aspect} to
the trajectory.

It may be noted that any member of the ensemble of trajectories satisfies
the required PDF $\vert\psi (x)\vert^{2}$. Eq. (4) depends upon the value of 
$dx/dt$ which, according to Eq. (6), is {\it independent} of the unknowable $%
t_{0}$.

\section{CASE OF A PARTICLE IN A BOX}

To find a particular trajectory we have to specify a particular case $\psi
(x)$. Consider the case where the particle is in a box of length $L$. With
the origin for coordinate $x$ at the center of the box, the wave function in
the ground state obeys [3]

\begin{equation}
\psi (x)=\sqrt{2/L}\cos (\pi x/L),\,\,-L/2\leq x\leq L/2.  \label{(8)}
\end{equation}

Using this in Eq. (5) gives a solution

\begin{equation}
x/L+(2\pi )^{-1}\sin (2\pi x/L)=(t-t_{0})/T.  \label{(9)}
\end{equation}
This defines the trajectory $x(t)$ of the particle for positions $x$ over
the interval $(0,L)$. Since the left-hand side of Eq. (9) is transcendental
in $x$, the equation must be inverted numerically for the desired trajectory 
$x(t)$.  A particle that traverses the given curve $x(t) $ back and forth
any integral number of times will satisfy the required PDF $(2/L)\cos
^{2}(\pi x/L).$ As a check on the theory, the $x$-values along the curve can
be `binned' to form a histogram $p(x)$ of occurrences $x$.  One would then
ascertain that it matches the theoretical answer $(2/L)\cos ^{2}(\pi x/L)$
to the accuracy defined by the finite bin size We have done this taking $%
\Delta x= 0.02 $. The results are not displayed for the sake of saving space.

\section{NON-STATIONARY PROBLEMS}

We may now generalize the approach. Suppose that the potential $V$ is now
more generally a function of both $x$ and $t$. Then the SWE solution is of a
generally non-separable form $\Psi (x,t)$. Accordingly, the particle has a
PDF on position that varies with the time,

\begin{equation}
|\Psi (x,t)|^{2}=p(x|t).  \label{(10)}
\end{equation}
The vertical solidus on the right-hand side emphasizes that this PDF on
random values $x$ is at (conditional upon) a fixed, nonrandom value $t$ of
time. May the particle still be describable by a trajectory $x(t)$?

The random variable in the amplitude function $\Psi (x,t)$ is $x$. The time $%
t$ is assumed known. However, in practice, a particle is tracked with
uniform randomness (1) in time. The net PDF on $x$ over all such time values
obeys the partition law of statistics [2],

\begin{equation}
p_{X}(x)=\int dt\,p(x|t)p_{T}(t).  \label{(11)}
\end{equation}
Then by Eqs. (1) and (10), Eq. (11) becomes

\begin{equation}
p_{X}(x)=\frac{1}{T}\int dt\,|\Psi (x,t)|^{2}.  \label{(12)}
\end{equation}
This is now used in place of quantity $|\psi (x)|^{2}$ in Eq. (5) to get the
required trajectory. If the positions $x$ along this trajectory are sampled
with uniform randomness in time, the resulting histogram will obey $p_{X}(x)$
given by Eq. (12).

\section{MOMENTUM AND PHASE VALUES}

We now turn to the question of the particle's momentum values $\mu $ ({\em %
note}: this is not the usual notation $p,$ because of previous use of $p$ to
denote probabilities.) The momentum of a purely classical particle obeys $%
\mu =mv=m\,dx/dt$, suggesting that a momentum `trajectory' $\mu (t)$ could
be found by simply differentiating $d/dt$ the previously found trajectory $%
x(t)$. However, in general this gives an ill-defined answer for the
corresponding PDF on momentum (or velocity $v$), as shown next.
Corresponding to Eq. (2) is

\begin{equation}
p_{V}(v)|dv|=p_{T}(t)|dt|,or,\,p_{V}(v)=\frac{1}{T|dv/dt|},  \label{(13)}
\end{equation}
after use of Eq. (1). Next, by Eq. (4)

\begin{equation}
\frac{dv}{dt}=\frac{d}{dt}(\frac{dx}{dt})=\frac{d}{dt}(\frac{1}{T|\psi
(x)|^{2}})=-\frac{2}{T|\psi (x)|^{3}}\frac{d|\psi |}{dx}v,  \label{(14)}
\end{equation}
after use of the chain rule of differentiation. Substitution in Eq. (13)
gives

\begin{equation}
p_{V}(v)=\frac{|\psi |^{3}}{2|v(d|\psi |/dx)|}.  \label{(15)}
\end{equation}
\qquad Although values of $\psi (x)$ are continuous, its derivatives $d\psi
/dx$ can be non-continuous [3]. At such velocities a PDF $p_{V}(v)$ computed
via Eq. (15) would be ill-defined. Hence, this approach to finding $p_{V}(v)$
is untenable in general. \ Likewise the corresponding PDF on momentum by
Jacobian transformation to $\mu =mv$ would be ill-defined in general. \
Hence, the momentum of our particle cannot be obtained as the mass times
velocity. \ This leaves the momentum representation of Heisenberg ($\mu
\rightarrow -i\,\hbar \,grad$) as the alternative. Thus, the probability
amplitude $\Phi (\mu )$ on momentum values $\mu $ is the usual Fourier
transform of $\Psi (x,t)$. This Fourier relation implies that the particle
obeys the Heisenberg uncertainty principle, as it must.

In summary, this approach to particle dynamics defines the particle
positions by the stochastic process $x(t;t_{0})$, where trajectory $x(t)$ is
deterministic up to the unknown constant $t_{0}$, and defines momentum
values by working in the usual Heisenberg representation $%
momentum\,\,\rightarrow -i\, \hbar \,grad$.

A general wave amplitude $\psi (x)$ is complex and hence has a finite phase
function. How is phase information reflected in the {\it trajectory }{\em %
x(t)}? The answer is that it is NOT, since by Eq. (4) $dx/dt$ is blind to
the phase function. Instead, the phase function follows from the Heisenberg
representation for momentum, by taking in the usual way the inverse
Fourier-transform of the probability amplitude $\Phi (\mu )$ on momentum.

\section{Effective potential $\overline{V}$}

%\vskip 3mm
The trajectory {\em x(t), }\ which we constructed to obey the SWE, can also
be made to obey Newton's laws, through the effect of an appropriate
potential. \ Newton's second law is

\begin{equation}
F=-\frac{\partial \overline{V}}{\partial x}=m\frac{dv}{dt},  \label{(16a)}
\end{equation}
\ where $\overline{V}$ is to be an effective potential function. The
acceleration $dv/dt$ borne of the SWE obeys Eq. (14). \ Using this in Eq. (%
\ref{(16a)}) forces the acceleration to be responding as well to a classical
force {\em F}. \ Using as well Eq. (4) for {\em v}, gives

\begin{equation}
-\frac{d\overline{V}}{dx}=-\frac{2m}{T^{2}|\psi |^{5}}\frac{d|\psi |}{dx}%
,\,\ \psi =\psi (x).
\end{equation}
This equation defines the effective potential function $\overline{V}$ $(x)\,$%
that must be present for the particle {\em to simultaneously} obey the SWE
and Newton's second law. \ It may be integrated directly, to give

\begin{equation}
\overline{V}(x)=-\frac{m}{2T^{2}}|\psi (x)|^{-4} .  \label{(18)}
\end{equation}

\ The dependence upon $\psi $ makes sense classically in that where the
particle tends not to be, i.e. at {\em x} for which $\psi (x)$ is low, the
potential tends to be high, acting as a {\em classical barrier}. \ Also, the
negative sign means that the effective potential is always attractive, i.e.,
tending to produce periodic motion.

In the more general case (Sec. IV) of a time-dependent potential {\em %
V(x,t), }by similar steps the effective potential obeys

\begin{equation}
\overline{V}(x)=-\frac{m}{2T^{2}}p_{X}(x)^{-2}.
\end{equation}

\section{Discussion}

The semi-classical theory of D. Bohm [1] attempts, as here, to ascribe a
classical trajectory to a particle that {\it a priori} obeys the SWE. As
with our approach, the particle trajectory by the Bohm approach is only
knowable to an unknown, additive constant. However, Bohm's theory differs
from ours both operationally (mathematically) and in its physical
assumptions. Operationally, instead of Eq. (4), the Bohm trajectory for a
particle obeys [4]

\begin{equation}
\frac{dx}{dt}=\frac{j(x,t)}{|\psi (x,t)|^{2}},  \label{(16)}
\end{equation}
where $j(x,t)$ is the quantum mechanical probability current. Comparing Eqs.
(4) and (\ref{(16)}) shows that, unless the current is unity, {\it the two
approaches give different trajectories for the particle}. In general the
current obeys

\begin{equation}
j(x,t)=\frac{i\hbar }{2m}[\Psi \frac{\partial \Psi ^{\ast }}{\partial x}%
-\Psi ^{\ast }\frac{\partial \Psi }{\partial x}].  \label{(17)}
\end{equation}
This rarely has the value unity. For example, if $\Psi $ is of the separable
form $\Psi (x,t)=(1/L)^{-1/2}\,\,\exp (ikx)\,\exp (-iEt/\hbar )$, Eq. (\ref
{(17)}) gives a current $j=\frac{\hbar k}{mL}$. Or, where $\Psi $ is of the
separable form $\Psi (x,t)=\psi (x)\exp (iEt/\hbar )$ with $\psi (x)$ real,
\ Eq. (\ref{(17)}) gives a current $j=0$. Hence the Bohm trajectories
usually differ from ours. \ As an example, in the particle-in-box case
above, $j=0$ so that by Eq. (\ref{(16)}) the Bohm trajectory obeys {\em dx/dt%
} = 0, i.e., the Bohm particle does not move in the box. As with our
effective potential Eq. (\ref{(18)}), the Bohm particle moves in the field
of an effective potential function that depends upon the wave amplitude
function. \ \ However the Bohm effective potential, commonly called the
''quantum potential'', is of an entirely different form [1].

The physical underpinning of the Bohm approach consists of four assumptions.
\ We list these in the following, along with corresponding assumptions of
our particle model:

(1) A SWE solution $\Psi (x,t)$ is not a probability amplitude but, rather,
a ``field'' analogous to the electromagnetic \ \ \ \ \ \ field. \ Our
approach preserves the standard probabilistic nature of $\Psi (x,t)$ . \ 

(2) The phase $S(x,t)$ of $\Psi (x,t)$ is also a classical Hamilton-Jacobi
function of mechanics. \ We do not assume this. \ Instead, the classical
trajectory follows from the special nature of the time coordinate as in \
Eqs. (1) and (2). \ 

(3) The Bohm semi-classical particle is acted upon by both the given
potential function $V(x,t)$ and {\em an added }potential function called the
``quantum potential'' where the latter is a function of $\Psi (x,t)$ as
well. \ Our classical particle is acted upon {\em in parallel }by the two
potential functions of the problem. \ These are (i) a conventional potential
function {\em V(x)} that defines, via the SWE, the probability field $|\psi
(x)|^{2}$ on position {\em x }that the particle is required to obey, and
(ii) an effective potential function $\overline{V}(x)$ that constrains the
classical particle to take the {\em particular trajectory} that satisfies
the required $|\psi (x)|^{2}$.\ \ \ The effective potential function $%
\overline{V}(x)$ derives from {\em the conventional} potential function {\em %
V(x) }via solution $|\psi (x)|^{2}$ to the SWE and Eq. (\ref{(18)}). \ In
this manner the SWE and the conventional potential function {\em V(x)} fix
both the statistics of the particle and its required trajectory (to an
additive constant). \ 

(4) The Bohm trajectory obeys Eq. (\ref{(16)}), which is based upon the
assumption that the quantum mechanical current {\em j} is also a classical
current. \ As we found, this assumption in general implies trajectories that
are different from ours. \ The assumption that the quantum mechanical
current {\em j} is a classical current is difficult to verify, since the
quantum mechanical current {\em j} is not a physical observable [3]. \ 

\vskip4mm

%\bigskip

{\bf REFERENCES}

1. D. Bohm, Phys. Rev. {\bf {85,} 166 (1952); P. R. Holland, {\it The
quantum theory of motion} (Cambridge University Press, Cambridge, England,
1993). }

{\bf 2. B.R.Frieden, {\em Probability, Statistical Optics and Data Testing,
2nd. ed.} (Springer-Verlag, New York, 1991) }.

{\bf 3. L.I. Schiff, {\em Quantum Mechanics} (McGraw-Hill, New York, 1955) }.

{\bf 4. D.Z. Albert,{\em \ Quantum Mechanics and Experience} (Harvard Univ.
Press, 1992); also see ``Bohm's alternative to quantum mechanics'', {\em %
Scientific American, } May, 1994, 58-67 for other references to Bohm's work. %
\vskip 4mm}

%\bigskip

\end{document}